\documentclass[letterpaper, 10 pt, conference]{ieeeconf}  

\IEEEoverridecommandlockouts                              

\usepackage{bigints}
\usepackage{soul}
\usepackage{epsfig} 
\usepackage{times} 
\usepackage{amsfonts,amssymb}
\usepackage{bm}
\usepackage{algorithm}
\usepackage{algorithmic}
\usepackage{float}
\usepackage{longtable,tabularx}

\usepackage{accents}

\usepackage{arydshln}
\usepackage{mwe}
\usepackage{graphbox} 
\usepackage{graphicx}
\usepackage{amsmath}
\usepackage[version=4]{mhchem}
\usepackage{siunitx}
\usepackage{longtable,tabularx}
\usepackage{booktabs}
\newtheorem{thm}{Theorem}[section]
\newtheorem{lem}[thm]{Lemma}

\newtheorem{defn}{Definition}[section]

\usepackage{multirow}
\usepackage{adjustbox}
\usepackage{xcolor}

\title{\LARGE \bf
Distributed $H_{\infty}$ Edge Weight Synthesis for Cooperative Systems
}

\author{Baris Taner$^{1}$ and Kamesh Subbarao$^{2}$
\thanks{*Support Number}
\thanks{$^{1}$Baris Taner is with Mechanical and Aerospace Engineering Department,
        University of Texas Arlington, 76019, Arlington, Texas, USA
        {\tt\small baris.taner@mavs.uta.edu}}%
\thanks{$^{2}$Dr. Kamesh Subbarao is with Mechanical and Aerospace Engineering Department,
        University of Texas Arlington, 76019, Arlington, Texas, USA
        {\tt\small kamesh.subbarao@uta.edu}}%
}

\begin{document}

\maketitle
\thispagestyle{empty}
\pagestyle{empty}

\begin{abstract}

This paper studies distributed edge weight synthesis of a cooperative system for a fixed topology to improve $H_{\infty}$ performance, considering that disturbances are injected at interconnection channels. This problem is cast into a linear matrix inequality problem by replacing original cooperative system with an equivalent ideal cooperative system. Derivations of the method relies on dissipative system framework. Proposed method provides an upper bound for the induced $\mathcal{L}_{2}$ norm of the original lumped cooperative system while reducing the computation time. A comparison for computation time illustrates the advantage of the proposed method against the lumped counterpart.

\end{abstract}

\section{INTRODUCTION}

Synthesis and control of cooperative systems (CSs) are active areas of research due to the attractive applications such as formation flight \cite{YU2020118}, satellite clustering \cite{Bai2018Satel} and sensor fusion \cite{Muller2016Sensor}. Some of these applications are clustering a few agents and form a small CS, while the others make large-scale CS, which might be geographically separated and life-critical \cite{Ding2019Survey}. Conventional way to approach these applications relies on lumped methods, where CS is modelled as a single system. However, this approach impose a heavy computational load in synthesis process even for relatively small CSs, which motivates the distributed  methodologies. 

There is a vast body of work in the area of distributed control and decision making for CSs. Some of the recent literature from different frameworks can be given as follows. In \cite{Ding2018DistOptCont}, an optimal control framework is used to solve multi-objective optimization problem for DC micro-grid using local-neighbour information. In another work, distributed impedance control is synthesized for event-triggered cooperative manipulation under disturbances using Lyapunov stability theory \cite{Dohman2020DistCont}. Finally, a distributed robust controller synthesis methodology from a dissipation perspective is utilized in \cite{weiland2016rob} to consider uncertainties with CS. The common basis for these methods is that, CS is modelled in distributed fashion.

In literature, synthesis of edge weights and topology of a CS is cast as a semi-definite programming (SDP). Following works can be given as examples, which are minimizing effective resistance of CSs using optimal allocation of the edge weights \cite{gosh2008} and obtaining better dynamic properties for CSs by meeting upper and lower bound constraints on maximum and Fiedler eigenvalues \cite{arcak2010}. Lumped problems for large scale CS can also be effectively solved using sparsity promoting methods. In a recent work, topology design of an undirected network is posed as an optimal control problem and solved using proximal gradient and Newton methods \cite{hasan2018TopSyn}. Similar examples can be given from distributed optimization of network systems, where cost function is separable into smaller convex cost functions \cite{yuan2016PDSub,HUANG2020105827}. All these works share a common basis, which is the lumped modelling and distributed optimization of CS.

\textit{Motivation.}To the best of authors knowledge, edge weight and topology synthesis of CSs are only posed using lumped CSs with distributed optimization methods in the literature. Yet this problem can be modelled in a distributed fashion before referring to distributed optimization methods. By this way, potentially faster converging algorithms will be available. Method originally defined in \cite{langbort2004Dist} and recently implemented for controller synthesis in \cite{weiland2016rob}, imposes a constraint on the interconnections by assuming the interconnections are ideal. This assumption results in a symmetric adjacency matrix, which is a limitation on the synthesis. This method is utilized in this paper for distributed edge weight synthesis, where performance is measured in terms of induced $\mathcal{L}_{2}$ norm of the system. Without the relaxations, this problem is NP-hard \cite{babazadeh2015nphard}. Edge synthesis in this work is done on a CS, whose individual agents are single input single output (SISO) stable systems. It should be noted that topology of the CS is predefined and not synthesized.  

\textit{Contributions.} This work resolves limitations on adjacency matrices and complexity due to NP-hard definition of the problem by redefining the CS. This is done by embedding adjacency matrix into a synthetic system definition and making ideal interconnections between original agents of the CS and the synthetic one. This results in a unique communication topology, which allows distributed CS modelling for any adjacency matrix. In addition, when underlying structure is exploited, this method eliminates the conditions leading to NP-hard problem by promoting sparsity in defining interconnection constraints. Key contributions of this paper are as follows:
\begin{itemize}
\item Adopts distributed CS modelling into an edge weight synthesis problem by resolving its limitations on adjacency matrix. 
\item Resolves complexity due to NP-hardness by introducing an equivalent sparse and ideal CS definition to replace original CS. By this way, problem becomes linear matrix inequality (LMI) optimization problem, which can be solved using SDP solvers. 
\end{itemize}

Paper is organized such that section \ref{sec:back}, expresses underlying Graph structure, lumped CS modelling, $H_{\infty}$ nominal performance for lumped CS model, distributed CS modelling and required definitions and lemmas to solve LMIs. Following that, problem formulation is given in section \ref{sec:prob}. Main result is shared in section \ref{sec:main}. This is followed by section \ref{sec:res}, where numerical verifications of the method is given for four agent case in both lumped and distributed fashions for comparison. In addition to that, distributed method proposed in this paper is tested upto 10 vehicle case. Finally, conclusions are provided in section \ref{sec:conc}.

\section{PRELIMINARIES}
\label{sec:back}

\subsection{Notation}
Let $\mathbb{R}$, $\mathbb{N}$, $\mathbb{S}^{p}$ and $\mathbb{S}_{s}^{p}$ denotes real numbers, natural numbers, symmetric and skew symmetric matrices of size $p$, respectively. $M_{ij}$ represents element in the $i^{th}$ row $j^{th}$ column of matrix ${\bm M}$. Matrix inequality conditions are defined with $<$ and $>$, which stands for $\leq$ and $\geq$, respectively. A matrix ${\bm M}:=\text{diag}(\cdot)$ is a block matrix, where diagonal entries are the arguments of $\text{diag}$. A matrix ${\bm M}:=\text{col}(\cdot)$ is a block matrix, where vertical entries are the arguments of $\text{col}$. Similarly, ${\bm M}:=\text{row}(\cdot)$ is a block matrix, where vertical entries are the arguments of $\text{col}$. Finally, $\mathcal{F}_{L}$ and $\mathcal{F}_{U}$ stands for lower and upper linear fractional transformation (LFT). 

\subsection{Underlying Graph Structure}
The interaction/communication among agents in the cooperative systems are described by Graph $\mathcal{G}=(\mathcal{N},\mathcal{E})$, which consists of node set $\mathcal{N}$ and edge set $\mathcal{E}$ \cite{ANDERSON2019108538}. Edge set $\mathcal{E}\subset\mathcal{N} \times \mathcal{N}$ is given between nodes $i\in \mathcal{N}$ and $j \in \mathcal{N}$ such that $(j,i) \in  \mathcal{E}$ denotes node $i$ receives information from $j$. Adjacency matrix ${\bm {\Upsilon}}=[\upsilon_{ij}] \in \mathbb{R}^{N \times N} $ of $\mathcal{G}$ is composed of weighting scalars ${\upsilon}$, where ${\upsilon}$ quantifies the strength of the connection from node $j$ to node $i$. $N$ is the number agents in the cooperative system (CS). Formally, ${{\Upsilon}}_{ij}$ is described as in the following equation.

\begin{equation}
\label{eq1}
{\bm {\Upsilon}}_{ij} = \begin{cases}
{\upsilon}>0, & j \neq i,~~ (j,i) \in \mathcal{E} \\
{\upsilon}=0, & otherwise
\end{cases}
\end{equation}

\subsection{Construction of the Lumped CS}

Construction of CS is described starting from agents, which are denoted as $^{i}{\bm G}$ for $i=\left[ 1, \cdots,N \right]$. The state space definition of $^{i}{\bm G}$ is given as in \eqref{eq2}, where $^{i}{\bm x} \in \mathbb{R}^{n_{xi}}$, $^{i}{\bm w}_{1} \in \mathbb{R}^{n_{w1i}}$, $^{i}{\bm w} \in \mathbb{R}^{n_{wi}}$, $^{i}{\bm z}_{1} \in \mathbb{R}^{n_{z1i}}$ and $^{i}{\bm z} \in \mathbb{R}^{n_{zi}}$. Agents share output information over the input - output ports denoted as ${}^{i}{\bm w}~-~{}^{i}{\bm z}$ such that ${}^{i}{\bm w}={\bm {\Upsilon}}{}^{i}{\bm z}$. Signals over this port is denoted as spatial signals. Channel through ports ${}^{i}{\bm w}_{1}~-~{}^{i}{\bm z}_{1}$ is the performance channel.

\begin{equation}
\label{eq2}
\left[ \begin{array}{c}
^{i}\dot{\bm x} \\ \hline
^{i}{\bm y}_{1} \\
^{i}{\bm y}
\end{array} \right] ~=~ \left[\begin{array}{c|cc}
^{i}{\bm A} & ^{i}{\bm B}_{1} & ^{i}{\bm B} \\ \hline
^{i}{\bm C}_{1} & ^{i}{\bm D}_{1} & ^{i}{\bm E}_{1} \\
^{i}{\bm C} & ^{i}{\bm F}_{1} & {\bm 0}
\end{array} \right] \left[ \begin{array}{c}
^{i}{\bm x} \\ \hline
^{i}{\bm w}_{1} \\
^{i}{\bm w}
\end{array} \right]
\end{equation}

A system ${\bm S}$ is created that represents a group of agents, which is simply described as ${\bm S}~=~\text{diag}\{ ^{i}{\bm G} \}_{i=1}^{N}$ and given in \eqref{eq3}, where ${\bm x} \in \mathbb{R}^{n_{x}}$, ${\bm w}_{1} \in \mathbb{R}^{n_{w1}}$, ${\bm w} \in \mathbb{R}^{n_{w}}$, ${\bm z}_{1} \in \mathbb{R}^{n_{z1}}$ and ${\bm z} \in \mathbb{R}^{n_{z}}$. Here $n_x = N \cdot n_{xi}$ and rest of the signals are expanded likewise.

\begin{equation}
\label{eq3}
\left[ \begin{array}{c}
\dot{\bm x} \\ \hline
{\bm z}_{1} \\
{\bm z}
\end{array} \right]~=\left[ \begin{array}{c|cc}
{\bm A} & {\bm B}_{1} & {\bm B} \\ \hline
{\bm C}_{1} & {\bm D}_{1} & {\bm E}_{1} \\
{\bm C} & {\bm F}_{1} & {\bm 0}
\end{array} \right]
\left[ \begin{array}{c}
{\bm x} 	\\ \hline
{\bm w}_{1} \\
{\bm w}
\end{array} \right]
\end{equation}

A cooperative system, ${\bm H}$, can be constructed using ${\bm {\Upsilon}}$ to connect vehicles to each other as represented by lower LFT such that ${\bm H}=\mathcal{F}_{L}({\bm S},{\bm {\Upsilon}})$. The state space representation of ${\bm H}$ is provided in \eqref{eq4}. 

\begin{equation}
\label{eq4}
\begin{aligned}
\left[ \begin{array}{c|c} {\bm {\mathcal{A}}} & {\bm {\mathcal{B}}}_{1} \\ \hline {\bm {\mathcal{C}}}_{1} & {\bm {\mathcal{D}}}_{1} \end{array} \right]  =\left[ \begin{array}{c|c}
{\bm A}+{\bm B}{\bm {\Upsilon}}{\bm C} & {\bm B}_{1}+{\bm B}{\bm {\Upsilon}}{\bm F}_{1} \\ \hline
{\bm C}_{1}+{\bm E}_{1}{\bm {\Upsilon}}{\bm C} & {\bm D}_{1}+{\bm E}_{1}{\bm {\Upsilon}}{\bm F}_{1}
\end{array} \right]
\end{aligned}
\end{equation}

\subsection{Feasibility Condition for $H_{\infty}$ Performance}

Assuming system given in \eqref{eq4} for a given ${\bm {\Upsilon}}$ is well-posed and asymptotically stable, then induced $\mathcal{L}_{2}$ norm of the system is given by the scalar $\gamma$ and $\| {\bm H} \|_{2} < \gamma$ for all inputs ${}^{i}{\bm w}_{1} \in \ell_{2}^{n_{w1i}}$ \cite{weiland2016rob}. This equivalently means that matrix inequalities (MIs) in \eqref{eq5} holds true. Pre- and post-multiplying the second inequality in \eqref{eq5} with $\left[ \begin{array}{cc} {\bm x}^{T} & {\bm w}_{1}^{T} \end{array} \right]$ and $\left[ \begin{array}{cc} {\bm x}^{T} & {\bm w}_{1}^{T} \end{array} \right]^{T}$, yields $ \gamma^{2} \| {\bm w}_{1} \|_{2}^{2}-\| {\bm z}_{1} \|_{2}^{2}<{\bm 0} $.
\begin{equation}
\label{eq5}
\begin{aligned}
\mathcal{X} & ~>~ {\bm 0} \\
\left[ \begin{array}{c}
{\bm {\mathcal{X}}} \\ {\bm 0}
\end{array} \right]\left[\begin{array}{cc} {\bm {\mathcal{A}}} & {\bm {\mathcal{B}}}_{1} \end{array} \right]+\left[ \begin{array}{c}{\bm {\mathcal{A}}}^{T} \\ {\bm {\mathcal{B}}}_{1}^{T} \end{array} \right]\left[ \begin{array}{cc}
{\bm {\mathcal{X}}} & {\bm 0}
\end{array} \right] \\
 ~+~ \left[ \begin{array}{c}
{\bm {\mathcal{C}}}_{1}^{T} \\ {\bm {\mathcal{D}}}_{1}^{T}
\end{array} \right]\left[ \begin{array}{cc}
{\bm {\mathcal{C}}}_{1} & {\bm {\mathcal{D}}}_{1}
\end{array} \right]+\left[ \begin{array}{cc}{\bm 0} & {\bm 0} \\ {\bm 0} & -\gamma^{2}{\bm I} \end{array} \right] &~<~ {\bm 0}
\end{aligned}
\end{equation}

Second inequality given in \eqref{eq5} depends non-linearly on matrix variables ${\bm {\mathcal{X}}}$ and ${\bm {\Upsilon}}$. This non-linear matrix inequality falls into the class of bilinear matrix inequalities (BMIs). Linearization and solution to this BMI can be obtained by multiple methods, where one of them is non-linear transformation as in \cite{scherer1997linear} and another one is given as in \cite{wang2018seqlmi}. Drawback of these methods are computational burden to calculate feasible solutions to ${\bm {\mathcal{X}}}$ and ${\bm {\Upsilon}}$ as the number of agents and number of states of the agents grow. Number of unknowns to be solved for this set of MIs is given by $(N \cdot n_{xi}(1+N \cdot n_{xi})/2)+n_{\alpha}$, where $n_{\alpha} \in \mathbb{N}$ denotes the number of independent variables used to define adjacency matrix to be synthesized. As clearly seen, state size of each agent and number of agents in the system quadratically increases the number of unknowns to be solved.

\subsection{Construction of Distributed CS for Arbitrary Interconnections}

CS given in \eqref{eq4} can also be described in a distributed fashion and this is realized by imposing constraints on the spatial signal channel of the agents. Unlike in lumped CS, constraints cannot reveal themselves in the state space representation, however, they are introduced as supply functions to the storage function of the CS. This is done by introducing these constraints in the MIs and this is summarized from the literature in the following text.

Distributed CS model relies on an interconnection constraint on spatial signal ports of each agent. This constraint is characterized by a supply function given as in \eqref{eq7}. The relevance of the supply functions to synthesize edge weight becomes clearer with the following definitions.

\begin{defn}[Dissipativity \cite{simpson2016Dissi}] ${\bm W}$ being input space, ${\bm Z}$ being output space and ${\bm X}$ being state space, let ${P}~:~{\bm W} \times {\bm Z} \rightarrow \mathbb{R}$ be a supply function.  A system ${\bm G}$ with supply function ${P}$ is dissipative if there exists a non-negative storage function $V~:~{\bm X} \rightarrow\mathbb{R}$ for admissible trajectories of $w$, $z$ and $x$, such that
\begin{equation}
\label{eq7}
V(x(t_{0}))+\int_{t_{0}}^{t_{1}} {P}(w(t),z(t))dt \geq V(x(t_{1})),~~\forall ~ t_{0} \leq t_{1}
\end{equation}
\end{defn}

Considering the $H_{\infty}$ performance, a quadratic supply function is known to be given by bounded real lemma \cite{park2021BRL,weiland2016rob} that is given as in \eqref{eq8} for the system $^{i}{\bm G}$.

\begin{equation}
\label{eq8}
^{i}{\Psi}~=~\left[ \begin{array}{c}
^{i}{\bm z}_{1} \\ \hline ^{i}{\bm w}_{1}
\end{array} \right]^{T} \left[ \begin{array}{c|c}
-\frac{1}{\gamma^2}{\bm I} & {\bm 0} \\ \hline
{\bm 0} & {\bm I}
\end{array} \right] \left[ \begin{array}{c}
^{i}{\bm z}_{1} \\ \hline ^{i}{\bm w}_{1}
\end{array} \right]
\end{equation}

\begin{figure}[thpb]
 	\centering
	\framebox{\parbox{2in}{\center \includegraphics[height=100pt]{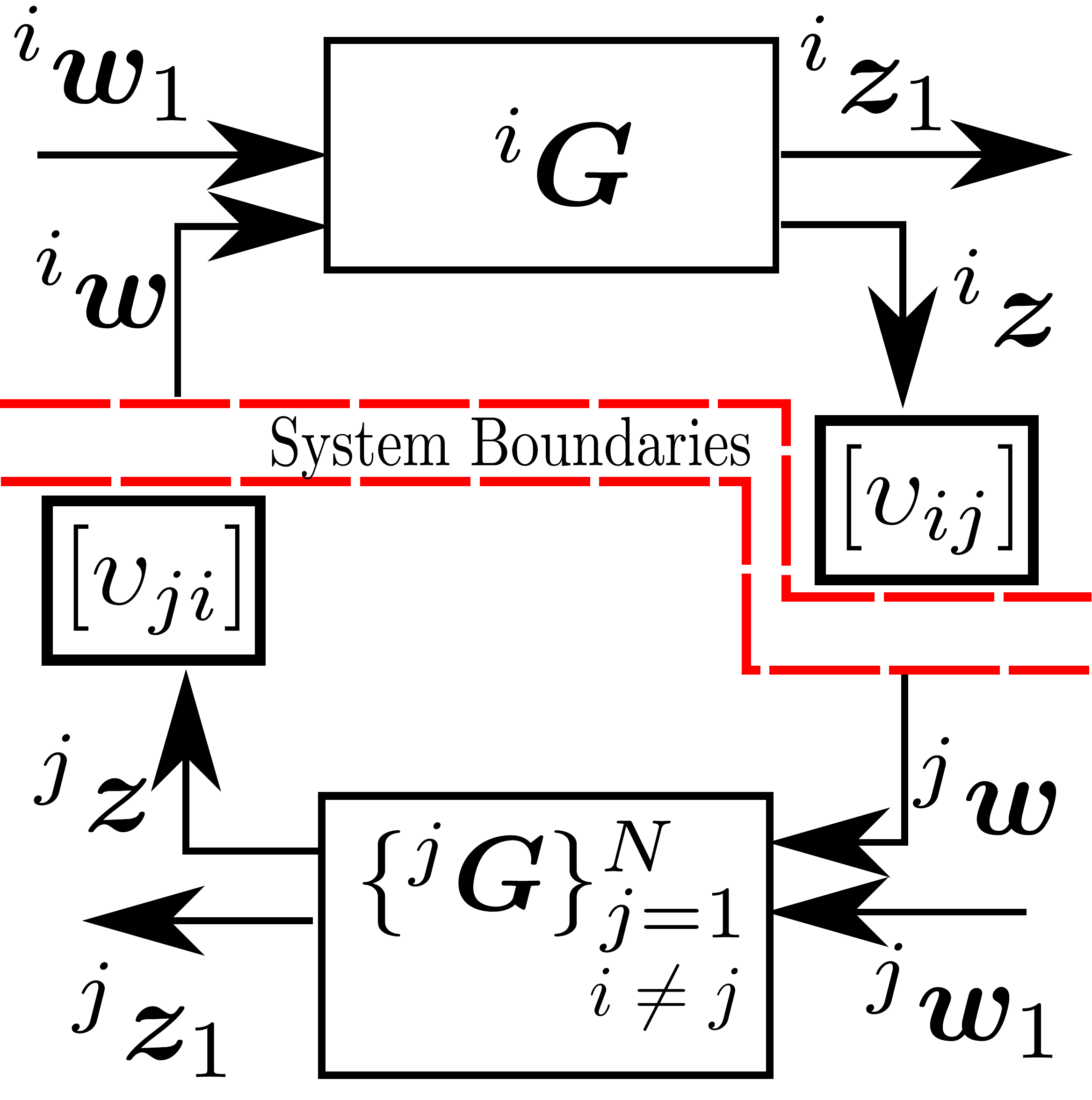}}}
	\caption{Interconnection of agents.}
\label{fig1}
\end{figure}

A general quadratic supply function can be given as in \eqref{eq9} to impose interconnections for a cooperative system by mapping spatial signals to agents as represented in \cite{langbort2004Dist,weiland2016rob}. Spatial signals are mapped within CS by ${\bm {\Upsilon}}$, and in \eqref{eq9}, ${\bm {\Upsilon}}$ is embedded in the correlation matrix, ${}^{i}{\bm Z} \in \mathbb{R}^{n_{zi}+n_{wi} \times n_{zi}+n_{wi}}$, which is partitioned as in \eqref{eq9}. This definition will be clarified as the paper progresses. For a CS, mapping of spatial signals is illustrated in Figure \ref{fig1}.

\begin{equation}
\label{eq9}
\begin{aligned}
^{i}{\Phi}&~=~\left[ \begin{array}{c}
^{i}{\bm z} \\ \hline ^{i}{\bm w}
\end{array} \right]^{T} \left[ \begin{array}{c|c}
^{i}{\bm Z}_{11} & ^{i}{\bm Z}_{12} \\ \hline
^{i}{\bm Z}_{12}^{T} & ^{i}{\bm Z}_{22}
\end{array} \right] \left[ \begin{array}{c}
^{i}{\bm z} \\ \hline ^{i}{\bm w}
\end{array} \right] \\
{}^{i}{\bm Z}&~=~\left[ \begin{array}{c|c} {}^{i}{\bm Z}_{11} & {}^{i}{\bm Z}_{12} \\ \hline {}^{i}{\bm Z}_{21} & {}^{i}{\bm Z}_{22} \end{array} \right]
\end{aligned}
\end{equation}

Considering the supply functions $^{i}{\Psi}$ and $^{i}{\Phi}$, a general supply function can be written as $^{i}{P}~=~{}^{i}{\Psi}+ {}^{i}{\Phi}$ and $H_{\infty}$ performance of the distributed CS can be checked using following MIs for $i=\left[ 1,\cdots,N \right]$.

\begin{equation}
\label{eq10}
\begin{aligned}
^{i}{\bm {{X}}} & ~>~ {\bm 0} \\
\left[ \begin{array}{ccc}^{i}{\bm {{X}}} ~ {}^{i}{\bm {{A}}} + {}^{i}{\bm {{A}}}^{T} ~ {}^{i}{\bm {{X}}} & ^{i}{\bm {{X}}} ~ {}^{i}{\bm {{B}}}_{1} & {}^{i}{\bm {{X}}}   ~ {}^{i}{\bm {{B}}} \\
{}^{i}{\bm {{B}}}_{1}^{T} ~ {}^{i}{\bm {{X}}} & -{\bm I} & {\bm 0} \\
{}^{i}{\bm {{B}}}^{T} ~ {}^{i}{\bm {{X}}} & {\bm 0} & {\bm 0}
\end{array} \right] & ~+~ \\ 
\frac{1}{\gamma^{2}} \left[ \begin{array}{c}
{}^{i}{\bm {{C}}}_{1}^{T} \\ {}^{i}{\bm {{D}}}_{1}^{T} \\ {}^{i}{\bm {{E}}}_{1}^{T}
\end{array} \right]^{T}
\left[ \begin{array}{ccc}
{}^{i}{\bm {{C}}}_{1} & {}^{i}{\bm {{D}}}_{1} & {}^{i}{\bm {{E}}}_{1}
\end{array} \right] &~+~ \\
\left[ \begin{array}{cc}
{}^{i}{\bm {{C}}}^{T} & {\bm 0} \\ {}^{i}{\bm {{F}}}_{1}^{T}  & {\bm 0} \\
{\bm 0} & {\bm I}
\end{array} \right]^{T} {}^{i}{\bm Z} \left[ \begin{array}{ccc}
{}^{i}{\bm {{C}}} & {}^{i}{\bm {{F}}}_{1} & {\bm 0} \\ {\bm 0} & {\bm 0} & {\bm I} \end{array} \right] &~<~ {\bm 0}
\end{aligned}
\end{equation}

\section{PROBLEM FORMULATION}
\label{sec:prob}

This section provides the framework to synthesize the edge weights denoted as ${\upsilon}$ in a distributed fashion using dissipative system framework.

Input and output sizes of each agent are modified to simplify the algebraic manipulations by introducing zeros to these vectors. New system is called as $^{i}\bar{\bm G} \in \mathbb{R}^{N \cdot (n_{z1i}+n_{zi}) \times N \cdot (n_{w1i}+n_{wi})}$. This extension is done as given in \eqref{eq6}, where $\star$ stands for ${\bm w}_{1}$, ${\bm w}$, ${\bm z}_{1}$ and ${\bm z}$. Simply, if $1^{st}$ agent is being modified then $^{i}\bar{\bm w}_{1}^{T}~=~\left[{\bm w}_{1}^{T},~{\bm 0}_{n_{w1i},1}^{T},~\hdots,~{\bm 0}_{n_{w1i},1}^{T} \right]^{T}$. Same formulation applies to all vectors. System matrices of $^{i}\bar{\bm G}$ is also extended to comply with the input-output sizes except for $^{i}\bar{\bm A}$, which is $^{i}\bar{\bm A}~=~^{i}{\bm A}$. The extended system matrices are denoted as $^{i}\bar{\bm B}_{1} \in \mathbb{R}^{n_{xi} \times N \cdot n_{w1i}}$, $^{i}\bar{\bm B} \in \mathbb{R}^{n_{xi} \times N \cdot n_{wi}}$, $^{i}\bar{\bm C}_{1} \in \mathbb{R}^{N \cdot n_{z1i} \times n_{xi}}$, $^{i}\bar{\bm C} \in \mathbb{R}^{N \cdot n_{zi} \times n_{xi}}$, $^{i}\bar{\bm D}_{1} \in \mathbb{R}^{N \cdot n_{z1i} \times N \cdot n_{w1i}}$, $^{i}\bar{\bm E}_{1} \in \mathbb{R}^{N \cdot n_{z1i} \times N \cdot n_{wi}}$ and $^{i}\bar{\bm F}_{1} \in \mathbb{R}^{N \cdot n_{zi} \times N \cdot n_{w1i}}$. Extension of these matrices are left to the reader for brevity of the paper.

\begin{equation}
\label{eq6}
\begin{aligned}
^{i}\bar{\bm \star}~=~&col\{ {\bm \star}_{j} \}_{j=1}^{N},~~~i=\left[ 1,\cdots,N\right] \\ 
{\bm \star}_{j}~=~& \begin{cases}
{\bm 0}, & j \neq i \\
{\bm \star}, & j = i
\end{cases}
\end{aligned}
\end{equation}

For the extended system, dissipativity is verified using supply function $^{i}{P}$ yet with a modification in ${}^{i}{\phi}$. A matrix is defined as $^{i}\tilde{\bm Z} \in \mathbb{R}^{(n_{zi}+n_{wi})\cdot N \times (n_{zi}+n_{wi})\cdot N}$, which shares the same partition with ${}^{i}{\bm Z}$ as given in \eqref{eq9}. $^{i}\tilde{\bm Z}$ reveals the effect of edge weights as given in \eqref{eq12}. The logic behind ${}^{i}{\tilde{\bm Z}}$ is that, ${}^{i}{\bm z}$ is scaled by scalar ${\upsilon}_{ij}$ before transferred to agent $j$ for all the agents, which is illustrated in Figure \ref{fig1}. If this mapping is followed, ${}^{i}{\tilde{\bm Z}}$ constructs an interconnection constraint on inputs and outputs of extended system ${}^{i}{\bar{\bm G}}$. Using the definition given in \eqref{eq12} MIs given in \eqref{eq11} can be constructed.

\begin{equation}
\label{eq12}
\resizebox{.95\hsize}{!}{$
\begin{aligned}
^{i}\tilde{\bm Z}~=~ 
\left[ \begin{array}{cc}{\bm {\Upsilon}} & {\bm 0}\\ {\bm 0} & {\bm I} \end{array} \right]^{T} & \left[ \begin{array}{cc}
^{i}\bar{\bm Z}_{11} & ^{i}\bar{\bm Z}_{12} \\  
^{i}\bar{\bm Z}_{12}^{T} & ^{i}\bar{\bm Z}_{22}
\end{array} \right] \left[ \begin{array}{cc}{\bm {\Upsilon}} & {\bm 0}\\ {\bm 0} & {\bm I} \end{array} \right] 
\end{aligned} $}
\end{equation}

\begin{equation}
\label{eq11}
\begin{aligned}
^{i}{\bm {{X}}} & ~>~ {\bm 0} \\
\left[ \begin{array}{ccc}^{i}{\bm {{X}}} ~ {}^{i}\bar{\bm {{A}}} + {}^{i}\bar{\bm {{A}}}^{T} ~ {}^{i}{\bm {{X}}} & ^{i}{\bm {{X}}} ~ {}^{i}\bar{\bm {{B}}}_{1} & {}^{i}{\bm {{X}}}   ~ {}^{i}\bar{\bm {{B}}} \\
{}^{i}\bar{\bm {{B}}}_{1}^{T} ~ {}^{i}{\bm {{X}}} & -{\bm I} & {\bm 0} \\
{}^{i}\bar{\bm {{B}}}^{T} ~ {}^{i}{\bm {{X}}} & {\bm 0} & {\bm 0}
\end{array} \right] ~+~ & \\ 
\frac{1}{{\gamma}^{2}} \left[ \begin{array}{c}
{}^{i}\bar{\bm {{C}}}_{1}^{T} \\ {}^{i}\bar{\bm {{D}}}_{1}^{T} \\ {}^{i}\bar{\bm {{E}}}_{1}^{T}
\end{array} \right]
\left[ \begin{array}{ccc}
{}^{i}\bar{\bm {{C}}}_{1} & {}^{i}\bar{\bm {{D}}}_{1} & {}^{i}\bar{\bm {{E}}}_{1}
\end{array} \right] ~+~ & \\
\left[ \begin{array}{cc}
{}^{i}\bar{\bm {{C}}}^{T} & {\bm 0} \\ {}^{i}\bar{\bm {{F}}}_{1}^{T} & {\bm 0} \\ {\bm 0}  & {\bm I}
\end{array} \right] {}^{i}\tilde{\bm Z} \left[ \begin{array}{ccc}
{}^{i}\bar{\bm {{C}}} & {}^{i}\bar{\bm {{F}}}_{1} & {\bm 0} \\ {\bm 0} & {\bm 0} & {\bm I} \end{array} \right] &~<~ {\bm 0} \\
\end{aligned}
\end{equation}

It is clear that MIs given in \eqref{eq10}, is not suitable for adjacency weight synthesis as components of ${\bm {\Upsilon}}$ does not appear explicitly. On the other hand, synthesis of ${\bm {\Upsilon}}$ using the MIs given in \eqref{eq11} requires us to solve a non-linear matrix inequality. In Section \ref{sec:main}, this problem will be given a solution. 

\section{MAIN RESULT}
\label{sec:main}

\begin{figure}[thpb]
 	\centering
	\framebox{\parbox{2in}{\center \includegraphics[height=100pt]{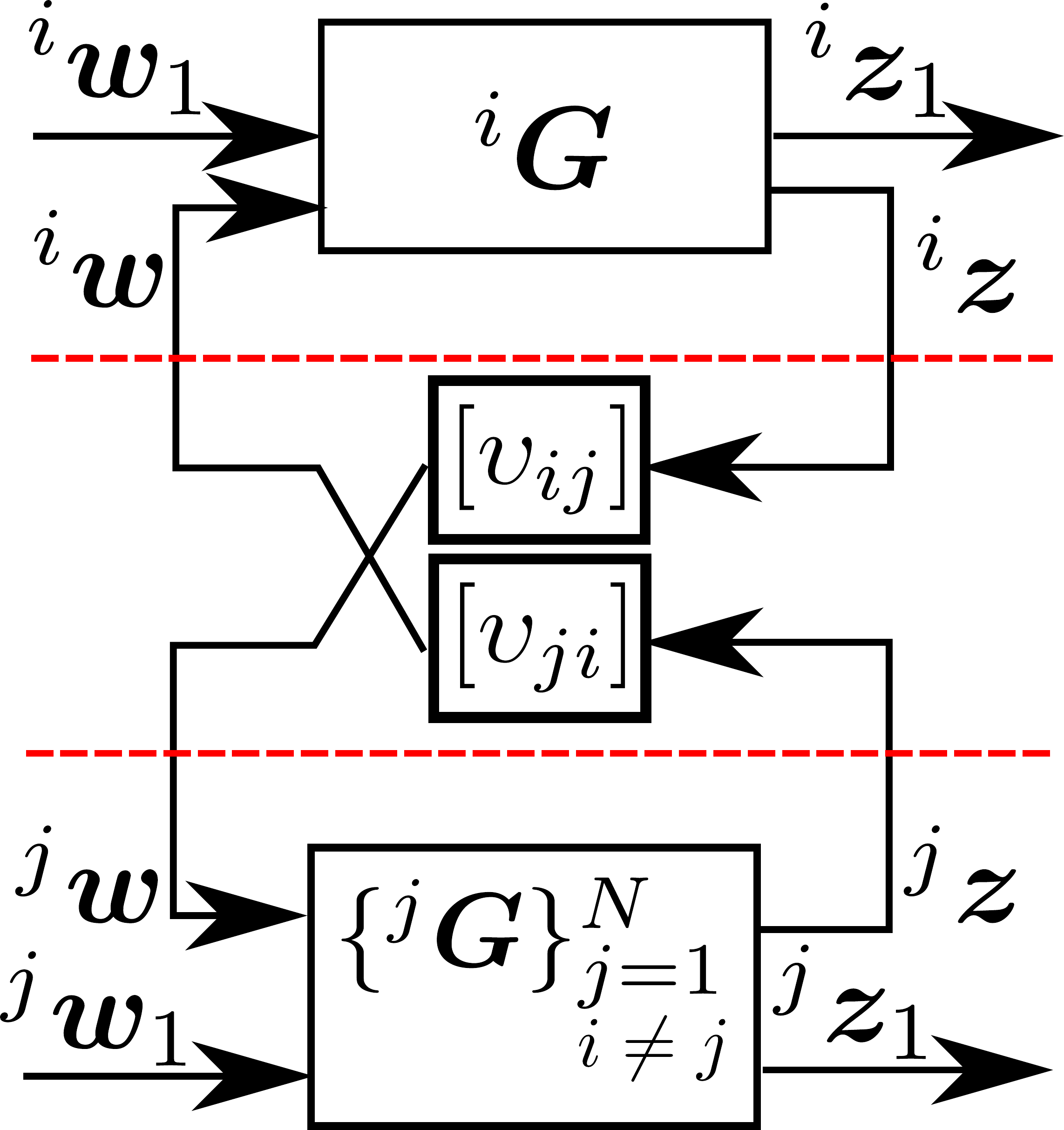}}}
	\caption{A new interpretation of the communication.}
\label{fig2}
\end{figure}

System given in Figure \ref{fig1}, is interpreted as $N$ agents sharing their information with a media that sinks the spatial signals and then sends them back again to the agents with some scaling. This media is specified with edge weights ${\upsilon}_{ij}$ and ${\upsilon}_{ji}$ as a group within the dashed red lines in Figure \ref{fig2}. Inputs and outputs of this group are ${}^{i}{z}$ and ${}^{i}{w}$, respectively.

This interpretation of the communication media allows us to assign a system representation to the communication itself, which is denoted in this paper as $^{c}{\bm G} \in \mathbb{R}^{n_{z} \times n_{w}}$. Representation of communication as a dynamic system is a concept used in dynamic Graph theory \cite{zecevic2010control}, however, this paper assumes that the edge weights are constants, for this reason $^{c}{\bm G}$ is given by a static feed-forward matrix as provided in \eqref{eq13}.

\begin{equation}
\label{eq13}
\begin{aligned}
{}^{c}{\bm G} := \left[ \begin{array}{c|c}
{\bm 0} & {\bm 0} \\ \hline
{\bm 0} & \Upsilon
\end{array} \right]
\end{aligned}
\end{equation}

\begin{figure}[thpb]
 	\centering
	\framebox{\parbox{2.4in}{\center \includegraphics[width=140pt]{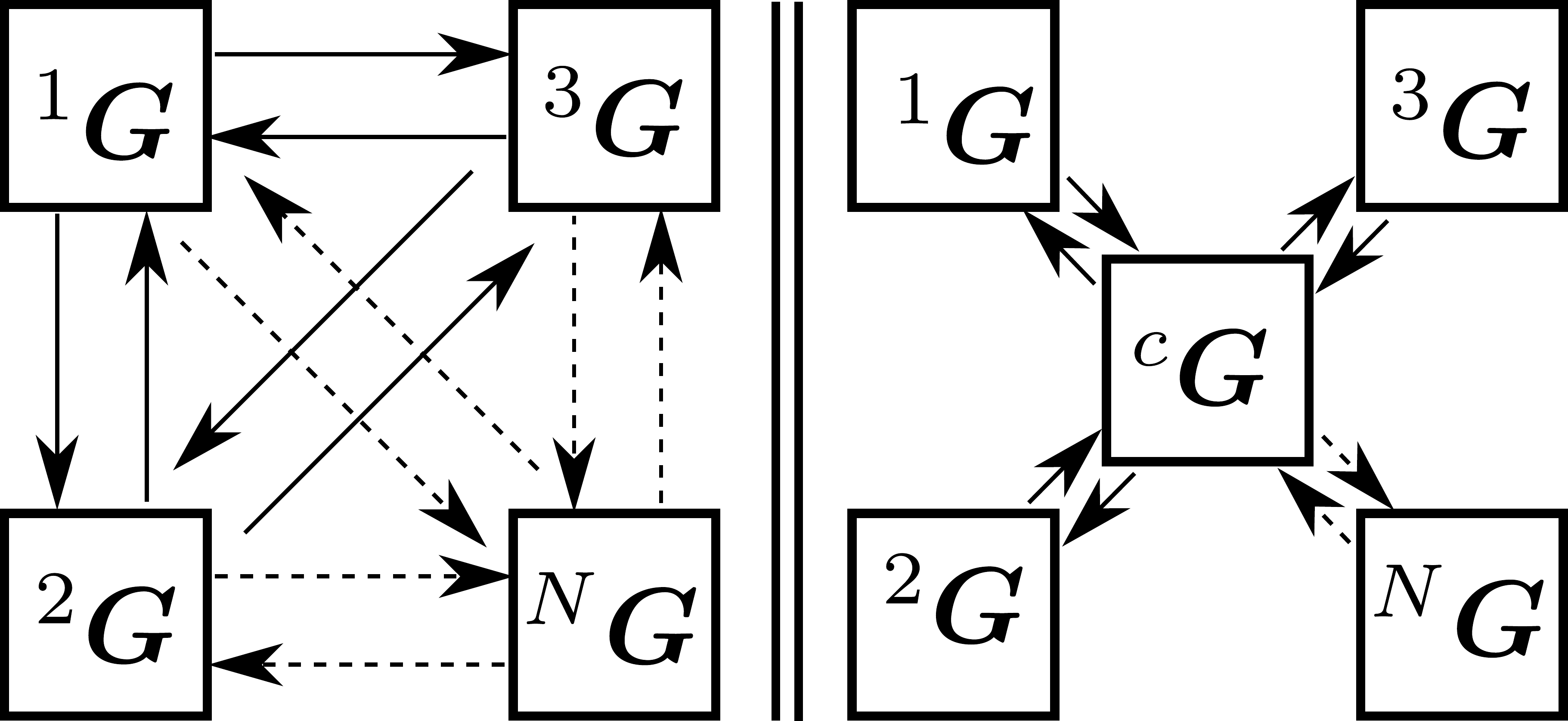}}}
	\caption{Conversion of the original CS (left) to the modified CS (right).}
\label{fig3}
\end{figure}

CS then takes a new form, which has ideal interconnections and these interconnections are between agents ${}^{i}{\bm G}$ and ${}^{c}{\bm G}$ only. The new interpretation of the CS is graphically given in Figure \ref{fig3}, where dashed lines show the interconnections that are being made between agents upto $N$.

Interconnections given in Figure \ref{fig3} is defined as in (\ref{eq14}), where ${}^{c}{\bm w}_{i} \in \mathbb{R}^{ n_{zi}}$ and ${}^{c}{\bm z}_{i} \in \mathbb{R}^{ n_{wi}}$ represents input to ${}^{c}{\bm G}$ from agent $i$ and output to agent $i$ from ${}^{c}{\bm G}$. 

\begin{equation}
\label{eq14}
\left[ \begin{array}{c}
{}^{i}{\bm z} \\
{}^{i}{\bm w}
\end{array} \right]~=~ \left[ \begin{array}{c}
{}^{c}{\bm w}_{i} \\
{}^{c}{\bm z}_{i}
\end{array} \right]~~~~\forall~i=\left[ 1, \cdots, N \right],~\forall~t \geq 0
\end{equation}

\begin{defn}[Neutrality of interconnections \cite{langbort2004Dist,weiland2016rob}]

Interconnections are considered to be neutral for neighbouring dissipative agents with respect to supply functions as defined in the form of \eqref{eq9}, if \eqref{eq14} is satisfied along with the condition in \eqref{eq15}, where ${}_{i}^{c}{\Phi}$ is the supply function of agent ${}^{c}{\bm G}$ and ${}_{c}^{i}{\Phi}$ is the supply function of agent ${}^{i}{\bm G}$ for $i=\left[ 1, \cdots, N \right]$.

\begin{equation}
\label{eq15}
\begin{aligned}
{}_{c}^{i}{\Phi}~+~{}_{i}^{c}{\Phi}~=~0, ~~~ \forall~i=\left[ 1, \cdots, N \right] \\
{}_{i}^{c}{\Phi}~=~\sum_{i=1}^{N} \left[ \begin{array}{c}
{}^{c}{\bm w}_{i} \\
{}^{c}{\bm z}_{i}
\end{array} \right]^{T}
\left[ \begin{array}{cc}
{}_{i}^{c}{\bm Y}_{11} & {}_{i}^{c}{\bm Y}_{12} \\
{}_{i}^{c}{\bm Y}_{21} & {}_{i}^{c}{\bm Y}_{22}
\end{array} \right]
\left[ \begin{array}{c}
{}^{c}{\bm w}_{i} \\
{}^{c}{\bm z}_{i}
\end{array} \right] \\ 
{}_{c}^{i}{\Phi}~=~\left[ \begin{array}{c}
{}^{i}{\bm w} \\
{}^{i}{\bm z}
\end{array} \right]^{T}
\left[ \begin{array}{cc}
{}^{i}{\bm Y}_{11} & {}^{i}{\bm Y}_{12} \\
{}^{i}{\bm Y}_{21} & {}^{i}{\bm Y}_{22}
\end{array} \right]
\left[ \begin{array}{c}
{}^{i}{\bm w} \\
{}^{i}{\bm z}
\end{array} \right]
\end{aligned}
\end{equation}
\end{defn}

Here a clarification should be made that, ${}^{c}{\bm G}$ makes interconnections with $N$ agents therefore it requires $N$ supply function definitions. ${}^{c}{\bm w}_{i}$ and ${}^{c}{\bm z}_{i}$ are the partitions of the input-output vectors ${}^{c}{\bm w}$ and ${}^{c}{\bm z}$ of ${}^{c}{\bm G}$ shared with ${}^{i}{\bm G}$. This dimensional complexity can be eliminated by applying the extension described in \eqref{eq6} on systems ${}^{i}{\bm G}$ and ${}^{c}{\bm G}$. With ${}^{c}{\bm G}$ there will be $N+1$ agents. Sizes of each system will be extended by $(N+1)N$ for the synthesis such that new systems are denoted as $^{i}\hat{\bm G} \in \mathbb{R}^{N \cdot (N+1) \cdot (n_{z1i}+n_{zi}) \times N \cdot (N+1) \cdot (n_{w1i}+n_{wi})}$. System matrices of $^{i}\hat{\bm G}$ is also extended to comply with the input-output sizes except for $^{i}\hat{\bm A}$, which is $^{i}\hat{\bm A}~=~^{i}{\bm A}$. The extended system matrices are denoted as $^{i}\hat{\bm B}_{1} \in \mathbb{R}^{n_{xi} \times N \cdot (N+1) \cdot n_{w1i}}$, $^{i}\hat{\bm B} \in \mathbb{R}^{n_{xi} \times N \cdot (N+1) \cdot n_{wi}}$, $^{i}\hat{\bm C}_{1} \in \mathbb{R}^{N \cdot (N+1) \cdot n_{z1i} \times n_{xi}}$, $^{i}\hat{\bm C} \in \mathbb{R}^{N \cdot (N+1) \cdot n_{zi} \times n_{xi}}$, $^{i}\hat{\bm D}_{1} \in \mathbb{R}^{N \cdot (N+1) \cdot n_{z1i} \times N \cdot (N+1) \cdot n_{w1i}}$, $^{i}\hat{\bm E}_{1} \in \mathbb{R}^{N \cdot (N+1) \cdot n_{z1i} \times N \cdot (N+1) \cdot n_{wi}}$ and $^{i}\hat{\bm F}_{1} \in \mathbb{R}^{N \cdot (N+1) \cdot n_{zi} \times N \cdot (N+1) \cdot n_{w1i}}$. Since $^{c}\hat{\bm G}$ does not have partitions like other agents do, it is convenient to write its system matrices as $^{c}\hat{\bm A}~=~{\bm 0} \in \mathbb{R}^{n_{xi}}$, $^{c}\hat{\bm B}~=~{\bm 0} \in \mathbb{R}^{n_{xi} \times N \cdot (N+1) \cdot (n_{w1i}+n_{wi})}$, $^{c}\hat{\bm C}~=~{\bm 0} \in \mathbb{R}^{N \cdot (N+1) \cdot (n_{z1i}+n_{zi}) \times n_{xi}}$ and $^{c}\hat{\bm D}~=~\text{diag}(\text{diag}({\bm 0},{\bm {\Upsilon}}),\text{diag}({\bm 0},{\bm {\Upsilon}})) \in \mathbb{R}^{N \cdot (N+1) \cdot (n_{z1i}+n_{zi}) \times N \cdot (N+1) \cdot (n_{w1i}+n_{wi})}$.

Relationship in \eqref{eq15}, has already been studied in the literature \cite{langbort2004Dist,weiland2016rob} and for completeness of the paper it will be shared here for this particular CS. From \eqref{eq14} and \eqref{eq15} one can set the algebraic relationship given in \eqref{eq16}.

\begin{equation}
\label{eq16}
\resizebox{.95\hsize}{!}{$
\begin{aligned}
\left[ \begin{array}{cc}
{}^{i}{\bm Y}_{11} & {}^{i}{\bm Y}_{12} \\
{}^{i}{\bm Y}_{21} & {}^{i}{\bm Y}_{22}
\end{array} \right] ~=~ - \left[ \begin{array}{cc}
{\bm 0} & {\bm I} \\
{\bm I} & {\bm 0}
\end{array} \right]& \left[ \begin{array}{cc}
{}_{i}^{c}{\bm Y}_{11} & {}_{i}^{c}{\bm Y}_{12} \\
{}_{i}^{c}{\bm Y}_{21} & {}_{i}^{c}{\bm Y}_{22}
\end{array} \right] \left[ \begin{array}{cc}
{\bm 0} & {\bm I} \\
{\bm I} & {\bm 0}
\end{array} \right]
\end{aligned} $}
\end{equation}
This relationship yields ${}^{i}{\bm Y}_{11}={}^{i}{\bm Y}_{11}^{T}=-{}_{i}^{c}{\bm Y}_{22}$ and ${}^{i}{\bm Y}_{12}^{T}=-{}_{i}^{c}{\bm Y}_{12}$ for $i=\left[1,\cdots,N \right]$.

Finally because of the communication topology for the modified system, a supply function for agents including $^{c}\hat{\bm G}$ can be formulated as in \eqref{eq18} with ${}^{d}\hat{\bm Z}$ for $d \in \{i,~c \}$. Partitioning of these matrices are given as ${}^{d}\hat{\bm Z}_{* \#}$, where $*,~\# \in \{ 1,~2 \}$ indicate the respective partition of ${}^{d}\hat{\bm Z}$. All zero matrices with respective sizes are defined by ${\bm 0}$.

\begin{equation}
\label{eq18}
\begin{aligned}
{}^{i}\hat{\bm Z}_{11}&~=~ \left[ \begin{array}{c|c}
\text{diag} \{ {\bm 0} \}_{i=1}^{N} & \text{col} \{ {\bm 0} \}_{i=1}^{N} \\ \hline
\text{row} \{ {\bm 0} \}_{i=1}^{N} & {}^{i}{\bm Y}_{11}
\end{array} \right] \\ 
{}^{i}\hat{\bm Z}_{12}&~=~ \left[ \begin{array}{c|c}
\text{diag} \{ {\bm 0} \}_{i=1}^{N} & \text{col} \{ {\bm 0} \}_{i=1}^{N} \\ \hline
\text{row} \{ {\bm 0} \}_{i=1}^{N} & {}_{i}^{c}{\bm Y}_{12}
\end{array} \right] \\ 
{}^{i}\hat{\bm Z}_{22}&~=~ \left[ \begin{array}{c|c}
\text{diag} \{ {\bm 0} \}_{i=1}^{N} & \text{col} \{ {\bm 0} \}_{i=1}^{N} \\
\hline
\text{row} \{ {\bm 0} \}_{i=1}^{N} & {}_{i}^{c}{\bm Y}_{11}
\end{array} \right] \\ 
{}^{c}\hat{\bm Z}_{11}&~=~ \left[ \begin{array}{c|c}
-\text{diag} \{ {}_{i}^{c}{\bm Y}_{11} \}_{i=1}^{N} & \text{col} \{ {\bm 0} \}_{i=1}^{N} \\ \hline
\text{row} \{ {\bm 0} \}_{i=1}^{N} & {\bm 0}
\end{array} \right] \\ 
{}^{c}\hat{\bm Z}_{12}&~=~ \left[ \begin{array}{c|c}
-\text{diag} \{ {}^{i}{\bm Y}_{12} \}_{i=1}^{N} & \text{col} \{ {\bm 0} \}_{i=1}^{N} \\ \hline
\text{row} \{ {\bm 0} \}_{i=1}^{N} & {\bm 0}
\end{array} \right] \\ 
{}^{c}\hat{\bm Z}_{22}&~=~ \left[ \begin{array}{c|c}
\text{diag} \{ {}^{i}{\bm Y}_{11} \}_{i=1}^{N} & \text{col} \{ {\bm 0} \}_{i=1}^{N} \\ \hline
\text{row} \{ {\bm 0} \}_{i=1}^{N} & {\bm 0}
\end{array} \right]
\end{aligned}
\end{equation}

Eventually, with supply functions of the form \eqref{eq8} and \eqref{eq9} described for the modified cooperative system following lemma can be given.

\begin{lem}
\label{lem1}
Let $^{i}\hat{\bm G}$ be a dissipative agent with state space realization is derived from \eqref{eq2} using the extension \eqref{eq6}. CS consisting of $^{i}\hat{\bm G}$ have a robust performance of $\gamma$ for input $^{i}\hat{w}_{1} \in \ell_{2}^{N \cdot (N+1)\cdot n_{w1i}}$ for $i=\left[1,\cdots,N \right]$ if there exists $^{i}{\bm X} \in \mathbb{S}^{n_{xi}}$, $^{i}{\bm Y}_{11} \in \mathbb{S}^{(N+1)\cdot n_{wi}}$, ${}_{i}^{c}{\bm Y}_{11} \in \mathbb{S}^{(N+1)\cdot n_{wi}}$, $^{i}{\bm Y}_{12} \in \mathbb{S}_{s}^{(N+1)\cdot n_{wi}}$ and ${}_{i}^{c}{\bm Y}_{12} \in \mathbb{S}_{s}^{(N+1)\cdot n_{wi}}$ such that

\begin{equation}
\label{eq19}
\begin{aligned}
^{i}\hat{\bm X} & ~\geq~ {\bm 0} \\
\left[ \begin{array}{ccc}^{i}\hat{\bm {{X}}} ~ {}^{i}\hat{\bm {{A}}} + {}^{i}\hat{\bm {{A}}}^{T} ~ {}^{i}\hat{\bm {{X}}} & ^{i}\hat{\bm {{X}}} ~ {}^{i}\hat{\bm {{B}}}_{1} & {}^{i}\hat{\bm {{X}}}   ~ {}^{i}\hat{\bm {{B}}} \\
{}^{i}\hat{\bm {{B}}}_{1}^{T} ~ {}^{i}\hat{\bm {{X}}} & -{\bm I} & {\bm 0} \\
{}^{i}\hat{\bm {{B}}}^{T} ~ {}^{i}\hat{\bm {{X}}} & {\bm 0} & {\bm 0}
\end{array} \right] ~+~ & \\ 
\frac{1}{\gamma^{2}} \left[ \begin{array}{c}
{}^{i}\hat{\bm {{C}}}_{1}^{T} \\ {}^{i}\hat{\bm {{D}}}_{1}^{T} \\ {}^{i}\hat{\bm {{E}}}_{1}^{T}
\end{array} \right]
\left[ \begin{array}{ccc}
{}^{i}\hat{\bm {{C}}}_{1} & {}^{i}\hat{\bm {{D}}}_{1} & {}^{i}\hat{\bm {{E}}}_{1}
\end{array} \right] ~+~ & \\
\left[ \begin{array}{cc}
{}^{i}\hat{\bm {{C}}}^{T} & {\bm 0} \\ {}^{i}\hat{\bm {{F}}}_{1}^{T} & {\bm 0} \\ {\bm 0}  & {\bm I}
\end{array} \right] {}^{i}\hat{\bm Z} \left[ \begin{array}{ccc}
{}^{i}\hat{\bm {{C}}} & {}^{i}\hat{\bm {{F}}}_{1} & {\bm 0} \\ {\bm 0} & {\bm 0} & {\bm I} \end{array} \right] &~<~ {\bm 0} \\
\left[ \begin{array}{cc}
{}^{c}\hat{\bm C}^{T} & {\bm 0} \\
{}^{c}\hat{\bm D}^{T} & {\bm I}
\end{array} \right]
\left[ \begin{array}{c|c}
\frac{1}{\gamma^{2}}{\bm I} & {\bm 0} \\ \hline
{\bm 0} & -{\bm I}
\end{array} \right]
\left[ \begin{array}{cc}
{}^{c}\hat{\bm C} & {}^{c}\hat{\bm D} \\ 
{\bm 0} & {\bm I}
\end{array} \right] ~+~ & \\
\left[ \begin{array}{cc}
{}^{c}\hat{\bm C}^{T} & {\bm 0} \\ 
{}^{c}\hat{\bm D}^{T} & {\bm I}
\end{array} \right] {}^{c}\hat{\bm Z} \left[ \begin{array}{cc}
{}^{c}\hat{\bm C} & {}^{c}\hat{\bm D} \\ 
{\bm 0} & {\bm I}
\end{array} \right] &~<~ {\bm 0}
\end{aligned}
\end{equation}
\end{lem}

It should be noted that, first two inequalities in \eqref{eq19} are LMIs.  Because of the specific structure of the interconnections of the third inequality ${}^{i}\hat{\bm Z}_{12}~=~{\bm 0}$. Eventually ${}^{i}\hat{\bm Z}$ and ${}^{c}\hat{\bm Z}$ are diagonal. By investigating \eqref{eq18} and  composite matrix $\left[ {}^{c}\hat{\bm C}~{}^{c}\hat{\bm D} \right]$ it can be deduced that
\begin{equation}
\label{eqaux19}
\left[ \begin{array}{cc}
{}^{c}\hat{\bm C} & {}^{c}\hat{\bm D}
\end{array} \right]^{T} \hat{\bm Z}_{11} \left[ \begin{array}{cc}
{}^{c}\hat{\bm C} & {}^{c}\hat{\bm D} \end{array} \right] ~=~ {\bm 0}
\end{equation}
As a result, third inequality in \eqref{eq19} is a quadratic MI. This inequality is linearised and solved about current solution as described in \textit{section 4.4} and \textit{section 4.5} of work \cite{wang2018seqlmi}. Proof of the lemma is given with the proof of Theorem \ref{thrm1}.

\begin{thm}\label{thrm1}
Let $^{i}\bar{\bm G}$ and $^{i}\hat{\bm G}$ be dissipative agents with state space realizations originates from \eqref{eq2} and extended as provided in \eqref{eq6}. The following two statements are valid:
\begin{enumerate}
\item CS consisting of $^{i}\bar{\bm G}$ achieves induced $\mathcal{L}_{2}$ gain performance of $\gamma^{*}$ for input $^{i}\bar{w}_{1} \in \ell_{2}^{N \cdot n_{w1i}}$ for $i=\left[1,\cdots,N \right]$ if there exist matrices ${}^{i}{\bm {{X}}}~=~{}^{i}{\bm {{X}}}^{T}$, ${}^{i}\bar{\bm Z}_{11}$, ${}^{i}\bar{\bm Z}_{12}$, ${}^{i}\bar{\bm Z}_{22}$ and $\Upsilon$ such that LMIs in \eqref{eq11} is satisfied for $i=\left[1,\cdots,N \right]$.

\item CS consisting of $^{i}\hat{\bm G}$ have a robust performance of $\gamma$ such that $\gamma^{*}<\gamma$ for input $^{i}\hat{w}_{1} \in \ell_{2}^{N \cdot (N+1)\cdot n_{w1i}}$ for $i=\left[1,\cdots,N \right]$
\end{enumerate}

\end{thm}

\textit{Proof.} As denoted previously, states of $^{i}{\bm G}$ is not modified during the extension to obtain $^{i}\bar{\bm G}$ and $^{i}\hat{\bm G}$. Extended input vectors are denoted as $\left[ {}^{i}\bar{\bm w}_{1}^{T}~{}^{i}\bar{\bm w}^{T} \right]^{T}$ and $\left[ {}^{i}\hat{\bm w}_{1}^{T}~{}^{i}\hat{\bm w}^{T} \right]^{T}$ for systems $^{i}\bar{\bm G}$ and $^{i}\hat{\bm G}$, respectively. This being said, second inequality in \eqref{eq11} is pre- and post-multiplied with state-input vector of $\left[ {}^{i}{\bm x}^{T}~{}^{i}\bar{\bm w}_{1}^{T}~{}^{i}\bar{\bm w}^{T} \right]^{T}$ regarding to the rules defined in extension given in \eqref{eq6}. This yields the following results.

\begin{equation}
\label{eq20}
\begin{aligned}
{}^{i}{\bm x}^{T}({}^{i}{\bm X}~{}^{i}{\bm A}+{}^{i}{\bm A}^{T}~{}^{i}{\bm X}){}^{i}{\bm x}~+~{}^{i}{\bm x}^{T}~{}^{i}{\bm X}~{}^{i}{\bm B}_{1}~{}^{i}{\bm w}_{1}~+~ \\
{}^{i}{\bm x}^{T}~{}^{i}{\bm X}~{}^{i}{\bm B}~{}^{i}{\bm w}~+~{}^{i}{\bm w}^{T}~{}^{i}{\bm B}^{T}~{}^{i}{\bm X}~{}^{i}{\bm x}~+~\\
{}^{i}{\bm w}_{1}^{T}~{}^{i}{\bm B}_{1}^{T}~{}^{i}{\bm X}~{}^{i}{\bm x}~-~{}^{i}{\bm w}_{1}^{T}{}^{i}{\bm w}_{1}~+~\frac{1}{\gamma^2}{}^{i}{\bm z}_{1}^{T}{}^{i}{\bm z}_{1}~+~\\
{}^{i}{\bm z}^{T}~{}^{i}\tilde{\bm Z}_{11}~{}^{i}{\bm z}~+~{}^{i}{\bm z}^{T}~{}^{i}\tilde{\bm Z}_{12}~{}^{i}{\bm w}~+~\\
{}^{i}{\bm w}^{T}~{}^{i}\tilde{\bm Z}_{21}~{}^{i}{\bm z}~+~{}^{i}{\bm w}^{T}~{}^{i}\tilde{\bm Z}_{22}~{}^{i}{\bm w}~\leq~0
\end{aligned}
\end{equation}

\begin{equation}
\label{eq21}
\begin{aligned}
{}^{i}\dot{\bar{V}}~=~& {}^{i}{\bm x}^{T}({}^{i}{\bm X}~{}^{i}{\bm A}+{}^{i}{\bm A}^{T}~{}^{i}{\bm X}){}^{i}{\bm x}~+~\\
&{}^{i}{\bm x}^{T}~{}^{i}{\bm X}~{}^{i}{\bm B}_{1}~{}^{i}{\bm w}_{1}~+~ {}^{i}{\bm x}^{T}~{}^{i}{\bm X}~{}^{i}{\bm B}~{}^{i}{\bm w}~+~\\
&{}^{i}{\bm w}^{T}~{}^{i}{\bm B}^{T}~{}^{i}{\bm X}~{}^{i}{\bm x}~+~{}^{i}{\bm w}_{1}^{T}~{}^{i}{\bm B}_{1}^{T}~{}^{i}{\bm X}~{}^{i}{\bm x} \\
{}^{i}\bar{\Phi}~=~&{}^{i}{\bm z}^{T}~{}^{i}\tilde{\bm Z}_{11}~{}^{i}{\bm z}~+~{}^{i}{\bm z}^{T}~{}^{i}\tilde{\bm Z}_{12}~{}^{i}{\bm w}~+~\\
& {}^{i}{\bm w}^{T}~{}^{i}\tilde{\bm Z}_{21}~{}^{i}{\bm z}~+~{}^{i}{\bm w}^{T}~{}^{i}\tilde{\bm Z}_{22}~{}^{i}{\bm w}
\end{aligned}
\end{equation}

Redefining \eqref{eq20} with ${}^{i}\dot{\bar{V}}$ and ${}^{i}\bar{\Phi}$ and summing over $i$ yields

\begin{equation}
\label{eq22}
\begin{aligned}
\sum_{i=1}^{N}{}^{i}\dot{\bar{V}}~+~\sum_{i=1}^{N}(\frac{1}{\gamma^2}{}^{i}{\bm z}_{1}^{T}{}^{i}{\bm z}_{1}~-~{}^{i}{\bm w}_{1}^{T}{}^{i}{\bm w}_{1})~+~\sum_{i=1}^{N}{}^{i}\bar{\Phi}~\leq~0.
\end{aligned}
\end{equation}

Integrating \eqref{eq22} along the admissible trajectory for $\left[ {}^{i}{\bm x}~{}^{i}\bar{\bm w}_{1}^{T}~{}^{i}\bar{\bm w}^{T} \right]^{T}$ from time $t=0$ to $t=T$ yields $\bar{V}( \{ {}^{i}{\bm x}(T) \}_{i=1}^{N}) - \bar{V}( \{ {}^{i}{\bm x}(0) \}_{i=1}^{N}) + \frac{1}{\gamma^2} \int_{0}^{T}\sum_{i=1}^{N}{}^{i}{\bm z}_{1}^{T}{}^{i}{\bm z}_{1}dt-\int_{0}^{T}\sum_{i=1}^{N}{}^{i}{\bm w}_{1}^{T}{}^{i}{\bm w}_{1}dt + \int_{0}^{T}\sum_{i=1}^{N}{}^{i}\bar{\Phi}dt ~\leq~0$. 

Following the same procedure for second inequality for $i=\left[1, \cdots ,N\right]$ and third inequality for $i~=~c$ of \eqref{eq19}, following results are obtained. This time column matrix that is being pre- and post-multiplied is $\left[ {}^{i}{\bm x}^{T}~{}^{i}\hat{\bm w}_{1}^{T}~{}^{i}\hat{\bm w}^{T} \right]^{T}$ recalling that rules of extension given in \eqref{eq6} applies.

\begin{equation}
\label{eq23}
\begin{aligned}
{}^{i}{\bm x}^{T}({}^{i}\hat{\bm X}~{}^{i}{\bm A}+{}^{i}{\bm A}^{T}~{}^{i}\hat{\bm X}){}^{i}{\bm x}~+~{}^{i}{\bm x}^{T}~{}^{i}\hat{\bm X}~{}^{i}{\bm B}_{1}~{}^{i}{\bm w}_{1}~+~ \\
{}^{i}{\bm x}^{T}~{}^{i}\hat{\bm X}~{}^{i}{\bm B}~{}^{i}{\bm w}~+~{}^{i}{\bm w}^{T}~{}^{i}{\bm B}^{T}~{}^{i}\hat{\bm X}~{}^{i}{\bm x}~+~\\
{}^{i}{\bm w}_{1}^{T}~{}^{i}{\bm B}_{1}^{T}~{}^{i}\hat{\bm X}~{}^{i}{\bm x}~-~{}^{i}{\bm w}_{1}^{T}{}^{i}{\bm w}_{1}~+~\frac{1}{\gamma^2}{}^{i}{\bm z}_{1}^{T}{}^{i}{\bm z}_{1}~+~\\
{}^{i}{\bm z}^{T}~{}^{i}\hat{\bm Z}_{11}~{}^{i}{\bm z}~+~{}^{i}{\bm z}^{T}~{}^{i}\hat{\bm Z}_{12}~{}^{i}{\bm w}~+~\\
{}^{i}{\bm w}^{T}~{}^{i}\hat{\bm Z}_{21}~{}^{i}{\bm z}~+~{}^{i}{\bm w}^{T}~{}^{i}\hat{\bm Z}_{22}~{}^{i}{\bm w}~+~ \\
~-~{}^{c}{\bm w}_{1}^{T}{}^{c}{\bm w}_{1}~+~\frac{1}{\gamma^2}{}^{c}{\bm z}_{1}^{T}{}^{c}{\bm z}_{1}~+~\\
{}^{c}{\bm z}^{T}~{}^{c}\hat{\bm Z}_{11}~{}^{c}{\bm z}~+~{}^{c}{\bm z}^{T}~{}^{c}\hat{\bm Z}_{12}~{}^{c}{\bm w}~+~\\
{}^{c}{\bm w}^{T}~{}^{c}\hat{\bm Z}_{21}~{}^{c}{\bm z}~+~{}^{c}{\bm w}^{T}~{}^{c}\hat{\bm Z}_{22}~{}^{c}{\bm w}~\leq~0
\end{aligned}
\end{equation}

\begin{equation}
\label{eq24}
\begin{aligned}
{}^{i}\dot{\hat{V}}~=~& {}^{i}{\bm x}^{T}({}^{i}\hat{\bm X}~{}^{i}{\bm A}+{}^{i}{\bm A}^{T}~{}^{i}\hat{\bm X}){}^{i}{\bm x}~+~\\
&{}^{i}{\bm x}^{T}~{}^{i}\hat{\bm X}~{}^{i}{\bm B}_{1}~{}^{i}{\bm w}_{1}~+~ {}^{i}{\bm x}^{T}~{}^{i}\hat{\bm X}~{}^{i}{\bm B}~{}^{i}{\bm w}~+~\\
&{}^{i}{\bm w}^{T}~{}^{i}{\bm B}^{T}~{}^{i}\hat{\bm X}~{}^{i}{\bm x}~+~{}^{i}{\bm w}_{1}^{T}~{}^{i}{\bm B}_{1}^{T}~{}^{i}\hat{\bm X}~{}^{i}{\bm x} \\
{}^{i}\hat{\Phi}~=~&{}^{i}{\bm z}^{T}~{}^{i}\hat{\bm Z}_{11}~{}^{i}{\bm z}~+~{}^{i}{\bm z}^{T}~{}^{i}\hat{\bm Z}_{12}~{}^{i}{\bm w}~+~\\
& {}^{i}{\bm w}^{T}~{}^{i}\hat{\bm Z}_{21}~{}^{i}{\bm z}~+~{}^{i}{\bm w}^{T}~{}^{i}\hat{\bm Z}_{22}~{}^{i}{\bm w} \\
{}^{c}\hat{\Phi}~=~&{}^{c}{\bm z}^{T}~{}^{c}\hat{\bm Z}_{11}~{}^{c}{\bm z}~+~{}^{c}{\bm z}^{T}~{}^{c}\hat{\bm Z}_{12}~{}^{c}{\bm w}~+~\\
& {}^{c}{\bm w}^{T}~{}^{c}\hat{\bm Z}_{21}~{}^{c}{\bm z}~+~{}^{c}{\bm w}^{T}~{}^{c}\hat{\bm Z}_{22}~{}^{c}{\bm w}
\end{aligned}
\end{equation}

\begin{equation}
\label{eq25}
\begin{aligned}
\sum_{i=1}^{N}{}^{i}\dot{\hat{V}}~+~\sum_{i=1}^{N}(\frac{1}{\gamma^2}{}^{i}{\bm z}_{1}^{T}{}^{i}{\bm z}_{1}~-~{}^{i}{\bm w}_{1}^{T}{}^{i}{\bm w}_{1})~+~ \\\sum_{i=1}^{N}{}^{i}\hat{\Phi}~+~{}^{c}\hat{\Phi}~+~(\frac{1}{\gamma^2}{}^{c}{\bm z}_{1}^{T}{}^{c}{\bm z}_{1}~-~{}^{c}{\bm w}_{1}^{T}{}^{c}{\bm w}_{1})~\leq~0
\end{aligned}
\end{equation}

$\sum_{i=1}^{N}{}^{i}\hat{\Phi}~+~{}^{c}\hat{\Phi}~=~0$ due to the fact that interconnections are neutral, which is dictated to the inequality by ${}^{i}\hat{\bm Z}$ and ${}^{c}\hat{\bm Z}$ matrices.  

\begin{equation}
\label{eq26}
\begin{aligned}
\sum_{i=1}^{N}{}^{i}\dot{\hat{V}}~+~\sum_{i=1}^{N}(\frac{1}{\gamma^2}{}^{i}{\bm z}_{1}^{T}{}^{i}{\bm z}_{1}~-~{}^{i}{\bm w}_{1}^{T}{}^{i}{\bm w}_{1})~+~ \\(\frac{1}{\gamma^2}{}^{c}{\bm z}_{1}^{T}{}^{c}{\bm z}_{1}~-~{}^{c}{\bm w}_{1}^{T}{}^{c}{\bm w}_{1})~\leq~0
\end{aligned}
\end{equation}

Integrating \eqref{eq26} along the admissible trajectory for $\left[ {}^{i}{\bm x}~{}^{i}\hat{\bm w}_{1}^{T}~{}^{i}\hat{\bm w}^{T} \right]^{T}$ from time $t=0$ to $t=T$ yields $\hat{V}( \{ {}^{i}{\bm x}(T) \}_{i=1}^{N}) - \hat{V}( \{ {}^{i}{\bm x}(0) \}_{i=1}^{N})  + \frac{1}{\gamma^2} \int_{0}^{T}\sum_{i=1}^{N}{}^{i}{\bm z}_{1}^{T}{}^{i}{\bm z}_{1}dt-\int_{0}^{T}\sum_{i=1}^{N}{}^{i}{\bm w}_{1}^{T}{}^{i}{\bm w}_{1}dt + \frac{1}{\gamma^2} \int_{0}^{T}{}^{c}{\bm z}_{1}^{T}{}^{c}{\bm z}_{1}dt-\int_{0}^{T}{}^{c}{\bm w}_{1}^{T}{}^{c}{\bm w}_{1}dt ~\leq~0$.

To prove $H_{\infty}$ performance, assume ${}^{i}{\bm x}(0)~=~{\bm 0}$ and ${}^{i}{\bm x}(T)~\neq~{\bm 0}$. Let ${}^{i}{\bm w}_{1}~\neq~{\bm 0}$ and ${}^{c}{\bm w}_{1}~\neq~{\bm 0}$ combined as ${\bm w}_{1}=\text{col}\{ \text{col}\{ {}^{i}{\bm w}_{1}\}_{i=1}^{N}, ~{}^{c}{\bm w}_{1} \}$. Let ${}^{i}{\bm z}_{1}$ and ${}^{c}{\bm z}_{1}$ combined as ${\bm z}_{1}=\text{col}\{ \text{col}\{ {}^{i}{\bm z}_{1}\}_{i=1}^{N}, ~{}^{c}{\bm z}_{1} \}$. Then from \eqref{eq26} we get
\begin{equation}
\label{eq27}
\frac{\|{\bm z}_{1}\|_{2}^{2}}{\| {\bm w}_{1} \|_{2}^{2}} \leq \gamma^{2}
\end{equation} 
for the CS an this completes the proof for Lemma \ref{lem1}. 

As seen in inequality \eqref{eq26}, modified CS definition replaces $\int_{0}^{T}\sum_{i=1}^{N}{}^{i}\bar{\Phi}dt$ with $\frac{1}{\gamma^2} \int_{0}^{T}{}^{c}{\bm z}_{1}^{T}{}^{c}{\bm z}_{1}dt-\int_{0}^{T}{}^{c}{\bm w}_{1}^{T}{}^{c}{\bm w}_{1}dt$, which is bounded quantity that depends on ${\bm {\Upsilon}}$. 

If one would seek neutrality condition \eqref{eq14} for the original CS (see Figure \ref{fig2}) and able to solve non-linear MI given in \eqref{eq11}, this means that each agent is dissipative. On top of that, due to \eqref{eq14}, \eqref{eqremark1} is true and this leads to \eqref{eqremark2}.
\begin{equation}
\label{eqremark1}
\int_{0}^{T}\sum_{i=1}^{N}{}^{i}\bar{\Phi}dt~=~0
\end{equation}

\begin{equation}
\label{eqremark2}
\begin{aligned}
\bar{V}( \{ {}^{i}{\bm x}(T) \}_{i=1}^{N}) - \bar{V}( \{ {}^{i}{\bm x}(0) \}_{i=1}^{N}) +& \\ \frac{1}{\gamma^2} \int_{0}^{T}\sum_{i=1}^{N}{}^{i}{\bm z}_{1}^{T}{}^{i}{\bm z}_{1}dt-
\int_{0}^{T}\sum_{i=1}^{N}{}^{i}{\bm w}_{1}^{T}{}^{i}{\bm w}_{1}dt &~\leq~0
\end{aligned}
\end{equation} 

Equation \eqref{eq26} infer that
\begin{equation}
\label{eqremark3}
\begin{aligned}
\hat{V}( \{ {}^{i}{\bm x}(T) \}_{i=1}^{N}) - \hat{V}( \{ {}^{i}{\bm x}(0) \}_{i=1}^{N})  + \\
\frac{1}{\gamma^2} \int_{0}^{T}\sum_{i=1}^{N}{}^{i}{\bm z}_{1}^{T}{}^{i}{\bm z}_{1}dt-\int_{0}^{T}\sum_{i=1}^{N}{}^{i}{\bm w}_{1}^{T}{}^{i}{\bm w}_{1}dt + \\
\frac{1}{\gamma^2} \int_{0}^{T}{}^{c}{\bm z}_{1}^{T}{}^{c}{\bm z}_{1}dt-\int_{0}^{T}{}^{c}{\bm w}_{1}^{T}{}^{c}{\bm w}_{1}dt ~\leq~0
\end{aligned}
\end{equation}

Let adjacency matrix be a weighted adjacency matrix and note that $\hat{V}( \{ {}^{i}{\bm x}(0) \}_{i=1}^{N})~=~\bar{V}( \{ {}^{i}{\bm x}(0) \}_{i=1}^{N})$ for ${}^{i}{\bm x}(0)~=~{\bm 0}$. Then from the subtraction of \eqref{eqremark2} from \eqref{eqremark3}, it can be calculated that $\hat{V}( \{ {}^{i}{\bm x}(T) \}_{i=1}^{N}) ~ - ~ \bar{V}( \{ {}^{i}{\bm x}(T) \}_{i=1}^{N})~\leq~\alpha$, where $0 ~ \leq ~ \alpha ~ \leq ~( \|{\bm {\Upsilon}}\|_{2} - 1 )\| {}^{c}{\bm w}_{1} \|_{2}^{2}$. Later inequality shows that, induced $\mathcal{L}_{2}$ norm of of $^{i}\hat{\bm G}$ is greater or equal than the one of $^{i}\bar{\bm G}$ and this completes the proof of Theorem \ref{thrm1}.

\section{NUMERICAL VERIFICATION}
\label{sec:res}

Methodology described under Section \ref{sec:main}, is verified using a CS of four vehicles. Adjacency matrix for lumped and distributed cases are compared in terms of edge weights ${\upsilon}$ and the robust performance level given by $\gamma$. Simulations are executed on a PC with Intel(R) Core(TM) i7-4720HQ CPU @2.6GHz, 16GB RAM running on Windows 10 OS and MATLAB 2019b. 

Dynamics of each agent is defined using state space representation given in \eqref{eq28}.

\begin{equation}
\label{eq28}
{}^{i}{\bm G}~=~\left[ \begin{array}{cc|c}
0 & 1 & 0 \\
-2 & -2 & 1 \\ \hline
1 & 0 & 0 
\end{array} \right]
\end{equation}

Agents are connected to each other with an adjacency matrix, ${\bm {\Upsilon}}$, which is defined as a function of a free scalar variable $\alpha$ and this is provided in \eqref{eq29}. $\alpha$ is contained in an interval defined as $\alpha~=~\left[0,~0.2 \right]$. This interval is selected arbitrarily without violating the fact that edge weight ${\upsilon}$ is $ 0 \leq {\upsilon} \leq 1$.

\begin{equation}
\label{eq29}
{\bm {\Upsilon}}~=~\left[\begin{array}{cccc}
0 & 2 \alpha & 3 \alpha & 1-5 \alpha \\
\alpha & 0 & 3 \alpha & 1-4 \alpha \\
\alpha & 2 \alpha & 0 & 1-3 \alpha \\
\alpha & 2 \alpha & 1-3 \alpha & 0
\end{array} \right]
\end{equation}

\begin{figure}[thpb]
 	\centering
	\framebox{\parbox{3in}{\center \includegraphics[width=200pt]{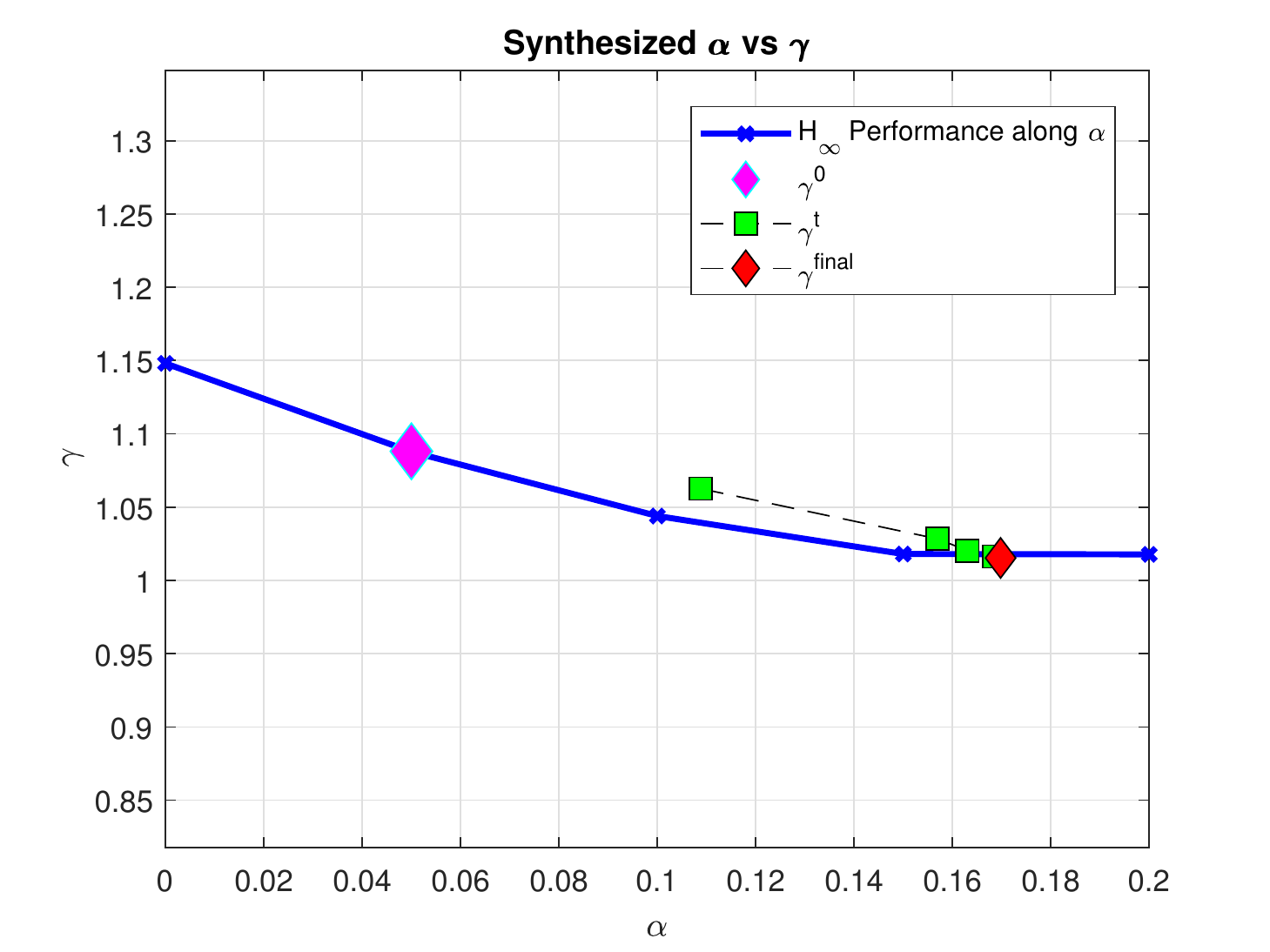}}}
	\caption{Synthesis results for lumped case}
\label{fig_res1}
\end{figure}

Using \eqref{eq28}, ${\bm S}$ is constructed for $N=4$. Then ${\bm H}$ is created as described in \eqref{eq4}. For this lumped model of the CS, $H_{\infty}$ performance is given in \eqref{eq5}, which is a BMI optimization problem and size of the decision variable is ${\bm {\mathcal{X}}} \in \mathbb{S}^{8}$. On top of that, there is $\alpha \in \mathbb{R}$. Thus, total number of variables to be solved for this case is $(8\cdot9)/2+1=37$. Although it is out of the scope of this paper, for proper comparison the method used to solve this problem should be noted. This problem is solved using sequential LMI method using convex-concave decompositions in 16.8125 s. Optimal values for $\alpha$ and $\gamma$ are given as $\alpha_{optimal}~=~0.1698$ and $\gamma_{optimal}~=~1.0153$ these values are provided on Figure \ref{fig_res1} with red diamond. Purple diamond represents the initial guess while green squares provided the calculated values for $\gamma$ and $\alpha$ in every iteration.

\begin{figure}[thpb]
 	\centering
	\framebox{\parbox{3in}{\center \includegraphics[width=200pt]{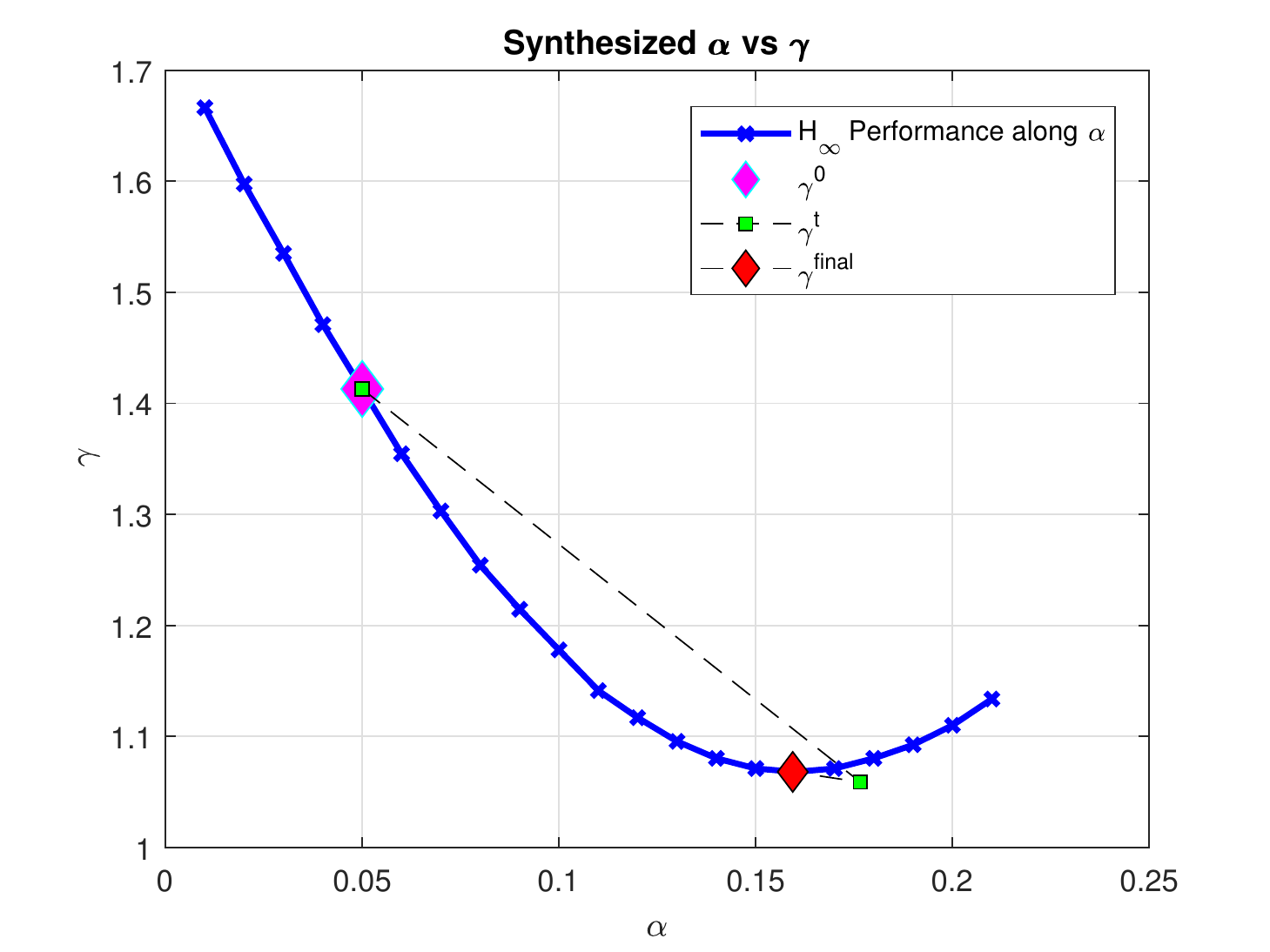}}}
	\caption{Synthesis results for distributed case.}
\label{fig_res2}
\end{figure}

For distributed case, agent dynamics is taken from the same system definition, which is given in \eqref{eq28}. Decision matrices denoted as ${}^{i}{\bm X} \in \mathbb{S}^{2}$ has $4\cdot (2\cdot 3)/2=12$ independent variables in total for $i=\left[ 1,\cdots,4\right]$. In addition to that, there are matrix variables ${}^{i}{\bm Y}_{11} \in \mathbb{S}^{4}$, which results in $((4\cdot 5)/2)\cdot 4 = 40$. Most of these variables are multiplied with zeros due to the extension given in \eqref{eq6} and disappear. Eventually, from each agent $4$ and in total $16$ independent variables are being calculated out of $40$. From  $^{c}\hat{\bm G}$, we have 16 variables du to ${}^{c}{\bm Y}_{11} \in \mathbb{S}^{4}$. Finally, $\alpha$ should be accounted in and this gives an overall $45$ variables for the whole simulation. The distributed edge weight synthesis problem is solved in 18.3750 s. Optimal values for $\alpha$ and $\gamma$ are given as $\alpha_{optimal}~=~0.1602$ and $\gamma_{optimal}~=~1.0712$ and these values are provided on Figure \ref{fig_res2} with red diamond. Initially, solver starts from the initial guess provided with the purple diamond, where $\alpha=0.05$. Compared to results of lumped method given in Figure \ref{fig_res1}, results of distributed case given in Figure \ref{fig_res2} show greater curvature and a greater infimum for the induced $\mathcal{L}_{2}$ norm for the CS.

\begin{figure}[thpb]
 	\centering
	\framebox{\parbox{2.75in}{\center \includegraphics[width=200pt]{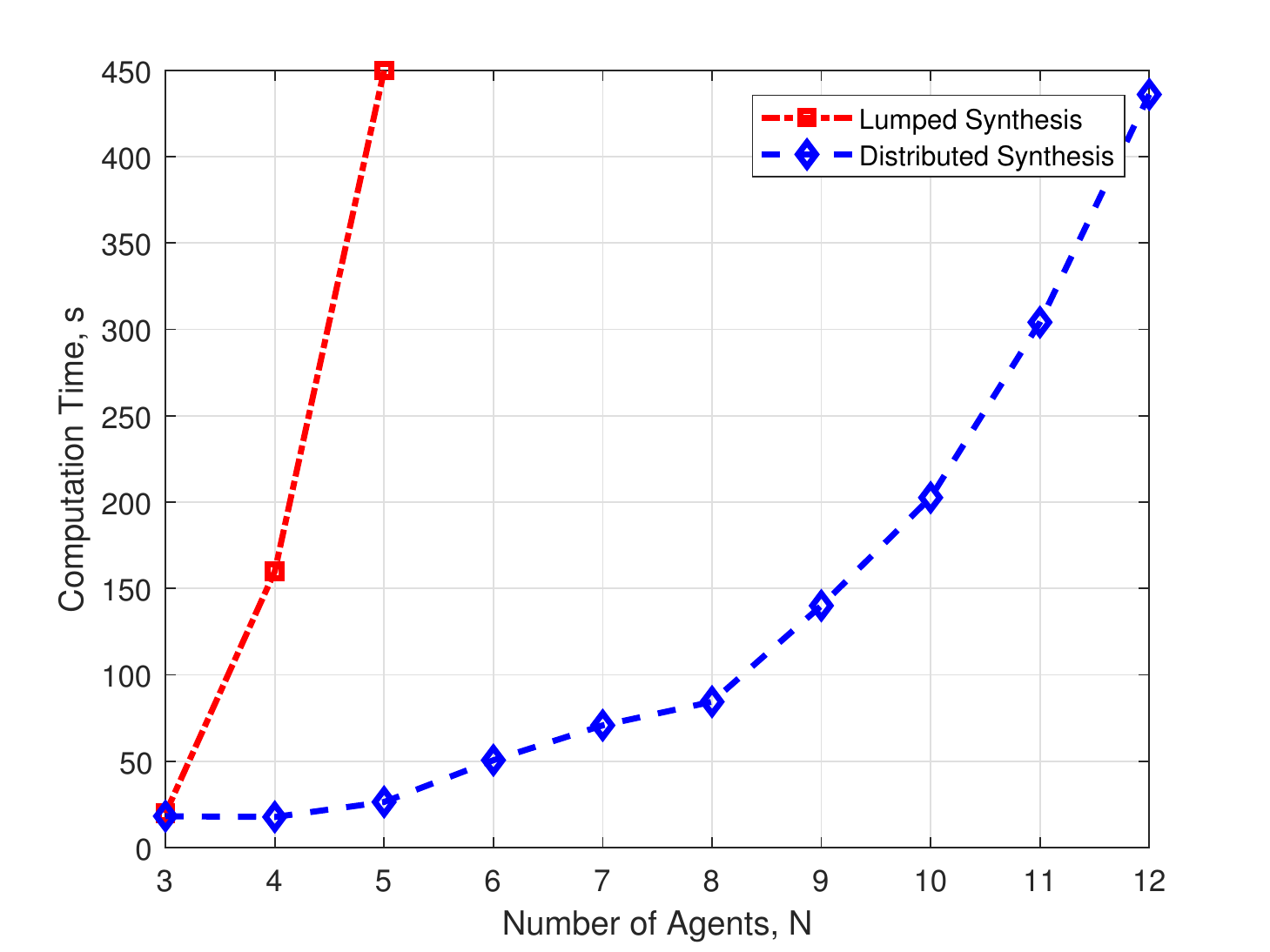}}}
	\caption{Computation time analysis of distributed and lumped methods.}
\label{fig_res3}
\end{figure}

Finally, a computational time analysis is executed for lumped and distributed cases, where number of agents for lumped and distributed cases are designated as $N_{lumped}=\left[ 3,~4,~5 \right]$ and $N_{dist}=\left[ 3,~ 4,~5,~6,~7,~8,~9,~10,~11,~12 \right]$. In both set of experiments, topology of the adjacency matrix is kept such that \eqref{eq29} for first four agents are fixed and each added agent listens to Agents $1$ and $N-1$. In $N=3$ case, first row and column is taken out. Under these conditions, Figure \ref{fig_res3} shows that computation time increases quadratically as the number of agents are increasing for the lumped case. Computation time shows quadratic behaviour for distributed case.

\section{CONCLUSIONS}
\label{sec:conc}

This paper focuses on finding the optimal values for edge weights in terms of enhancing $H_{\infty}$ performance criteria of a CS in a distributed fashion. Originally, this is an optimization problem with a BMI constraint, however, when the CS is modeled in a distributed fashion problem includes non-linear MIs as constraints. This problem is solved by modeling communication media as a synthetic agent denoted as ${}^{c}{\bm G}$ and making ideal interconnections between original agents and this synthetic agent. By this way, information on adjacency matrix is secured and distributed edge weight synthesis using dissipative system theory with neutrality condition on interconnections is executed.

Edge weight synthesis is illustrated on a cooperative system with four vehicles. Given the topology, edge weights of these systems are synthesized and results reveal following key conclusions.

\begin{enumerate}
\item Distributed method proposed in this paper is able to reach the optimal $\alpha$ value that is being calculated for lumped CS. The error between these values are 0.0099, which is 5.83\% error. Difference in curvature and infimum of the induced $\mathcal{L}_{2}$ norm is due to $\|{\bm {\Upsilon}}\|_{2}$ as explained in the proof of Theorem \ref{thrm1}.
\item In numerical verification, each agent has two states and this gives an advantage to lumped synthesis as there are 37 independent variables to be solved. In distributed case, there are 45 variables in total. It is clearly seen that increasing number of agents results in a quadratic increase in computation time for lumped case. The quadratic relationship can be given as $(n_{xi} \cdot N)^{2}$. On the other hand, number of variables to be solved in distributed case grow with a relationship of $n_{xi}^{2} \cdot {N} + n_{wi} \cdot N \cdot (N+1)$.
\item These relationships are visible in Figure~\ref{fig_res3} as complexity of the problem so the calculation time of the optimization problem increases quadratically for lumped method; However, computational time trend for distributed method shows a gradual shift from linear to quadratic as the number of agents increase. Dominance of these trends are decided by the size of state and spatial input/output vector of single agent. In addition to that number of states and number of spatial inputs/outputs to single agent define, when one method is superior to the other. Specifically, a larger spatial input/output vector for same state vector pushes the number of agents, when distributed method become superior, to higher values. It should also be noted that, there is a certain threshold for spatial input/output to state vector ratio that distributed method is no more superior for any number of agents.

\end{enumerate}

\addtolength{\textheight}{-12cm}   



%
%
\section*{ACKNOWLEDGMENT}

This work is supported by the Office of Naval Research via award number N00014-18-1-2215. The authors thank the anonymous reviewers for critically reading the manuscript and suggesting substantial improvements. This paper is accepted to CDC 2021 \cite{BarisSubbarao2021}.
%



\bibliographystyle{IEEEtran}
\bibliography{ieee_bib}

\end{document}